\documentclass[aps,prd,superscriptaddress,showpacs,preprintnumbers]{revtex4}
\usepackage{lineno,hyperref}
\usepackage{amssymb, graphics, setspace}
\usepackage{pgfplots}
\usepackage{textcomp}
\usepackage[caption=false]{subfig}
\usepackage{verbatim}
\usepackage{amsmath}
\usepackage{commath}

\usepackage{graphicx,bm}
\usepackage{url}
\usepackage{float}
\usepackage{textcomp}
\usepackage{verbatim}
\begin{document}

\newcommand{\dlt}{\bigtriangleup}
\newcommand{\beq}{\begin{equation}}
\newcommand{\eeq}[1]{\label{#1} \end{equation}}
\newcommand{\insertplot}[1]{\centerline{\psfig{figure={#1},width=14.5cm}}}

\parskip=0.3cm


\title{Approaching the asymptotics at the LHC}


\author{Norbert Bence}
\email{bencenorbert007@gmail.com}
\affiliation{Uzhgorod National University, \\14, Universytets'ka str.,  
	Uzhgorod, 88000, UKRAINE}

\author{L\'aszl\'o Jenkovszky}
\email{jenk@bitp.kiev.ua}
\affiliation{Bogolyubov Institute for Theoretical Physics (BITP),
	Ukrainian National Academy of Sciences \\14-b, Metrologicheskaya str.,
	Kiev, 03680, UKRAINE}

\author{Istv\'an Szanyi}
\email{sz.istvan03@gmail.com}
\affiliation{Uzhgorod National University, \\14, Universytets'ka str.,  
	Uzhgorod, 88000, UKRAINE}

\begin{abstract}
Recent results on the slope of the $pp$ diffraction cone measured by TOTEM at $7$ and $8$ GeV show an unexpected rapid rise in $s$, close to $B(s)\sim \ln^2$, rather than $\ln s$, typical of the Regge-pole predictions. We show that the new phenomenon can be accommodated by the inclusion of unitarity corrections to a simple Regge (pomeron) pole exchange.
Interestingly, the odderon may also promote the acceleration of $B(s)$. The onset of the new regime may be indicative of the approach to the asymptotic dynamics of strong interactions. We analyse the new data together with other available forward measurable in a unitarized Regge dipole. Unitarization proves crucial in fitting the data, especially those on the slope $B(s)$ showing a change from the $\ln (s)$ to $\ln^2 (s)$ behavior. Having fitted the free parameters of the unitarized model to the data, we predict the behavior of the cross sections and the slope at still higher energies, including those asymptotic.     
\end{abstract}


\maketitle

\section{Introduction} \label{s1}
The energy dependence of the forward slope of the diffraction cone is a basic observable related to the interaction radius. While in Regge-pole models the rise of the  total cross sections is regulated by the hardness of the Regge pole (here, the pomeron), the slope 
$B(s)$ in case of a single and simple Regge pole is always logarithmic. Deviation (acceleration) may arise from more complicated Regge singularities, the odderon (see Fig.~\ref{Fig:Bo}) and, definitely from unitarity corrections.

The TOTEM Collaboration announced \cite{Oster,Csorgo,Mario,Giani} the new results of the measurements on the proton-proton elastic slope at $7$ and $8$ TeV, $B$(7 TeV)$=19.84\pm0.27$ and $B$(8 TeV)$=19.9\pm0.3$ GeV$^{-2}$, which refers that its energy dependence is well beyond the value of the logarithmic approximation from lower energies, compatible with a $\ln s$ behavior (see slide 14 in \cite{Mario}). These data offer new information concerning the burning problem of the strong interaction dynamics, namely the onset of the approach to the asymptotic regime of the strong interaction. For similar studies see also \cite{TT}.  

The construction of any scattering theory consists of two stages: one first chooses an input amplitude ("Born term"), subjected to a subsequent unitarization procedure. Neither the input, nor the unitarization procedure are unique. In any case, the better the input, i.e. closer to the true amplitude, the better are the  chances of the unitarization. The standard procedure is that of Regge-eikonal, {\it i.e.} when the eikonal is identified with a simple Regge-pole input.

A possible alternative to the simple Regge-pole model as input is a double pole (double pomeron pole, or simply dipole pomeron, DP) in the angular momentum $(j)$ plane. It has a number of advantages over the simple pomeron Regge pole. In particular, it produces logarithmically risen cross sections already at the "Born" level, Sec. \ref{Sec:DP}. In this paper we analyze the forward slope of the diffraction cone. We start with a simple DP model and proceed with its unitarization. In doing so we use a unitarization procedure (Sec. \ref{Sec:Unitarity}), different from the familiar eikonalization \cite{Martin}.  

The approach to the expected asymptotic behavior has two stages. One is the onset pomeron dominance, i.e. domain where secondary reggeon contributions become negligible. It can be shown \cite{JLL} that in the nearly forward direction, at LHC energies the contribution from secondary trajectories is negligible, smaller than the error bars in the measured total cross section, {\it i.e.} "soft" physics at the LHC is pomeron-dominated. The next question is where does the pomeron itself reaches its asymptotics. Below we address these questions. 

The slope of the diffraction cone is defined as 
\begin{equation}\label{Eq:slope}
B(s,t\rightarrow 0)=\frac{d}{dt}\Biggl(\ln\frac{d\sigma}{dt}\Biggr)\bigg|_{t=0},
\end{equation}
where $A(s,t)$ is the elastic scattering amplitude.
In the case of a single and simple Regge pole, the slope increases logarithmically with $s$:
\begin{equation}\label{Eq:tslope}
B(s,t)=B_0(t)+2\alpha'(t)\ln(s/s_0).
\end{equation}
The slope of the trajectory is not constant, therefore the forward cone in the nearly forward direction deviating from the exponential, see \cite{Break} and earlier references therein.
The "forward" slope $B$ is extracted form the experimental data within finite bins in $t$ \cite{JL}, consequently it is function of both $s$ and $t$.

For example, at the SPS Collider the slope was measured at $\sqrt{s}=540$ GeV \cite{Burq} in two $t$-bins with the result
$$B=13.3\pm 1.5\ {\rm GeV}^{-2}\ \ \  (0.15<|t|<0.26)\ {\rm GeV}^2, $$
$$B=17.2\pm 1.0\ {\rm GeV}^{-2}\ \ \  (0.05<|t|<0.18)\ {\rm GeV}^2. $$ 
 
From these observations the following conclusions were drawn in Ref.\cite{JS}: 

1) the large values of B at the SPS may be reached only with a large, $>0.2$ GeV$^{-2}$ value of the pomeron slope $\alpha'$;

2) any analysis of the the slope $B$ must contain both $s$ and $t$ dependences.

Recall that at energies below the Tevatron, including those of SPS and definitely so at the ISR, secondary trajectories contribute significantly. In the unitarized version, Secs.~\ref{Sec:Unitarity}, \ref{Sec:Asympt}, for simplicity only the pomeron will be included. At the LHC and beyond, we are at the fortunate situation where these non-leading contribution may be neglected. Apart from the pomeron, the odderon may be present, although its contribution may be noticeable only away from the forward direction, {\it e.g.} at the dip \cite{JLL}. Recall also that a single Regge pole produces a monotonic $\ln s$ rise of the slope, discarded by the recent LHC data \cite{Mario}. 

The slope of the cone $B(s,t)$ is not measured directly, it is deduced from the data on the directly measurable differential cross sections within certain bins in $t$. Therefore, the primary sources are the cross sections or the scattering amplitude fitted to these cross sections. To scrutinize the slope we first prepare the ground by constructing (Subsection~\ref{ssec:model}) a model amplitude from which the slope can be calculated.
 
In Sec.~\ref{Sec:DP} we present an updated analysis of the main forward $pp$ observables
(elastic, inelastic and total cross sections, the slope $B$ as well as the ratio of real to imaginary part of the amplitude, based on the dipole pomeron (DP) model (for a review see {\it e.g.} \cite{PEPAN}). The DP pomeron model is unique alternative to a simple pole, higher order poles are not allowed by unitarity. Furthermore, the DP produces (logarithmically) rising cross sections even at unit intercept. The DP scales reproducing itself with respect to $s$-channel unitarity corrections. A particularly attractive feature of the DP is the built-in mechanism of the diffraction pattern: a single minimum appears, followed by a maximum in the differential cross section, confirmed by the experimental data in a wide span of energies. 

A unitarlization procedure is presented in Sec.~\ref{Sec:Unitarity}. The observable consequence of the unitarized DP, with emphasis on the LHC data, including those on the slope, are presented in Sec.~\ref{Sec:Unitarity}. These result not only secure unitarity but even provide an estimate of the approach to the asymptotic limit. Relevant estimates for the cross sections and other observables as well as their ratios are presented in Secs.~\ref{Sec:Unitarity} and \ref{Sec:Asympt}.

\section{The forward slope $B(s,t)$}\label{Sec:DP}
\subsection{DP fits to the low-$|t|$ data}\label{ssec:model}

In this subsection we prepare the ground by introducing the model amplitude and fitting its parameters to the data. At the LHC, the pomeron dominates, however, to be consistent with the lower energy data, particularly those from the ISR, we include also two secondary reggeones. The odderon, the pomeron's odd-$C$ counterpart may also be included in the fitting procedure (its identification requires the inclusion in the fits of $\bar pp$ data as well). however its contribution in the forward direction is negligibly small. 

The scattering amplitude is \cite{JLL}:

\begin{equation}\label{Eq:Amplitude}
A\left(s,t\right)_{pp}^{\bar pp}=A_P\left(s,t\right)+
A_f\left(s,t\right)\pm\left[A_{\omega}\left(s,t\right)+A_O\left(s,t\right)\right].
\end{equation}

Secondary reggeons are parametrized in a standard way \cite{KKL, KKL1}, with linear Regge trajectories and exponential residua. The $f$ and $\omega$ reggeons are the principal non-leading contributions to $pp$ or $\bar p p$ scattering:
\begin{equation}\label{Reggeon1}
A_f\left(s,t\right)=a_f{\rm e}^{-i\pi\alpha_f\left(t\right)/2}{\rm e}
^{b_ft}\Bigl(s/s_0\Bigr)^{\alpha_f\left(t\right)},
\end{equation}
\begin{equation}\label{Reggeon2}
A_\omega\left(s,t\right)=ia_\omega{\rm e}^{-i\pi\alpha_\omega\left(t\right)/2}{\rm e}
^{b_\omega t}\Bigl(s/s_0\Bigr)^{\alpha_\omega\left(t\right)},
\end{equation}
with $\alpha_f\left(t\right)=0.703+0.84t$ and
$\alpha_{\omega}\left(t\right)=0.435+0.93t$. 

As already mentioned, the pomeron is a dipole in the $j-$plane
\begin{equation}\label{Pomeron}
A_P(s,t)={d\over{d\alpha_P}}\Bigl[{\rm
	e}^{-i\pi\alpha_P/2}G(\alpha_P)\Bigl(s/s_{0P}\Bigr)^{\alpha_P}\Bigr]=
\end{equation}
$${\rm
	e}^{-i\pi\alpha_P(t)/2}\Bigl(s/s_{0P}\Bigr)^{\alpha_P(t)}\Bigl[G'(\alpha_P)+\Bigl(L-i\pi
/2\Bigr)G(\alpha_P)\Bigr].$$

Since the first term in squared brackets determines the shape of the cone, one fixes
\begin{equation} \label{residue} G'(\alpha_P)=-a_P{\rm
	e}^{b_P[\alpha_P-1]},\end{equation} where $G(\alpha_P)$ is recovered
by integration. Consequently the pomeron amplitude Eq.(\ref{Pomeron}) may be rewritten in the following ``geometrical'' form (for details see \cite{PEPAN} and references therein):
\begin{equation}\label{GP}
A_P(s,t)=i{a_P\ s\over{b_P\ s_{0P}}}[r_1^2(s){\rm e}^{r^2_1(s)[\alpha_P-1]}-\varepsilon_P r_2^2(s){\rm e}^{r^2_2(s)[\alpha_P-1]}],
\end{equation} where
$r_1^2(s)=b_P+L-i\pi/2,\ \ r_2^2(s)=L-i\pi/2,\ \ L\equiv
\ln(s/s_{0P})$ and the pomeron trajectory:

\begin{equation}\label{Ptray}
\alpha_P\equiv \alpha_P(t)= 1+\delta_P+\alpha_{1P}t - \alpha_{2P}(\sqrt{4m_{\pi}^2-t}-2m_{\pi}).
\end{equation}
The two-pion threshold in the trajectory accounts for the non-exponential behavior of the diffraction
cone at low $|t|$, see Ref.~\cite{Break}.

The odderon contribution is assumed to be of the same form as that of the pomeron apart from different values of adjustable parameters (labeled by the subscript ``$O$''): 
\begin{equation}\label{Odd}
A_O(s,t)={a_O\ s\over{b_O\ s_{0O}}}[r_{1O}^2(s){\rm
	e}^{r^2_{1O}(s)[\alpha_O-1]}   -\varepsilon_O r_2^2(s){\rm
	e}^{r^2_{20}(s)[\alpha_O-1]}],
\end{equation}
where
\begin{equation}
r_{1O}^2(s)=b_O+L-i\pi/2,\ \ r_{2O}^2(s)=L-i\pi/2,\ \ L\equiv
\ln(s/s_{0O}).
\end{equation}
and the trajectory

\begin{equation}\label{Eq:Otray}
\alpha_O\equiv \alpha_O(t) =
1+\delta_O+\alpha_{1O}t - \alpha_{2O}(\sqrt{4m_{\pi}^2-t}-2m_{\pi}).
\end{equation} 

In earlier versions of the DP the intercept of the pomeron was fixed at $\alpha(0)=1$, to avoid conflict with the Froissart bound. However later it was realized that the logarithmic rise of the total cross sections provided by the DP may not be sufficient to meet the data, therefore a supercritical intercept was allowed for. From the fits to the data the value $\epsilon=\alpha(0)-1=0.04,$ half of Landshoff's value \cite{Land} follows. This is understandable: the DP promotes half of the rising dynamics, thus moderating the departure from unitarity at the "Born" level (smaller unitarity corrections). Unitarization of the Regge-pole input is indispensable in any case, and in the next section we proceed along these lines.

We use the norm where
\begin{equation}\label{norm}
{d\sigma\over{dt}}(s,t)={\pi\over s^2}|A(s,t)|^2\ \  {\rm and}\ \
\sigma_{tot}(s)={4\pi\over s}\Im m A(s,t)\Bigl.\Bigr|_{t=0}\ .
\end{equation}

The free parameters of the model were simultaneously fitted to the data on elastic $pp$ and $p\bar p$ differential cross section in the region of the first cone, $|t|<1$ GeV$^2$ as well onto the data on total cross section and the ratio
\begin{equation}
\rho(s)=\frac{\Re e A(s,t=0)}{\Im m A(s,t=0)}
\end{equation}
in the energy range between $5$ GeV and $30000$ GeV. Our fitting strategy is to keep control of those parameters that govern the behavior of the forward slope, neglecting details that are irrelevant to the forward slope, such is the diffraction minimum (here related to absorption corrections through the parameter $\epsilon$ in Eq. (\ref{GP})). We find that the forward slope $B(s,0)$ is not affected by these corrections, therefore we set $\epsilon=0$. 

\begin{figure}[H] 
	\centering
	\subfloat[\label{fig:sigma}]{%
	\includegraphics[scale=0.19]{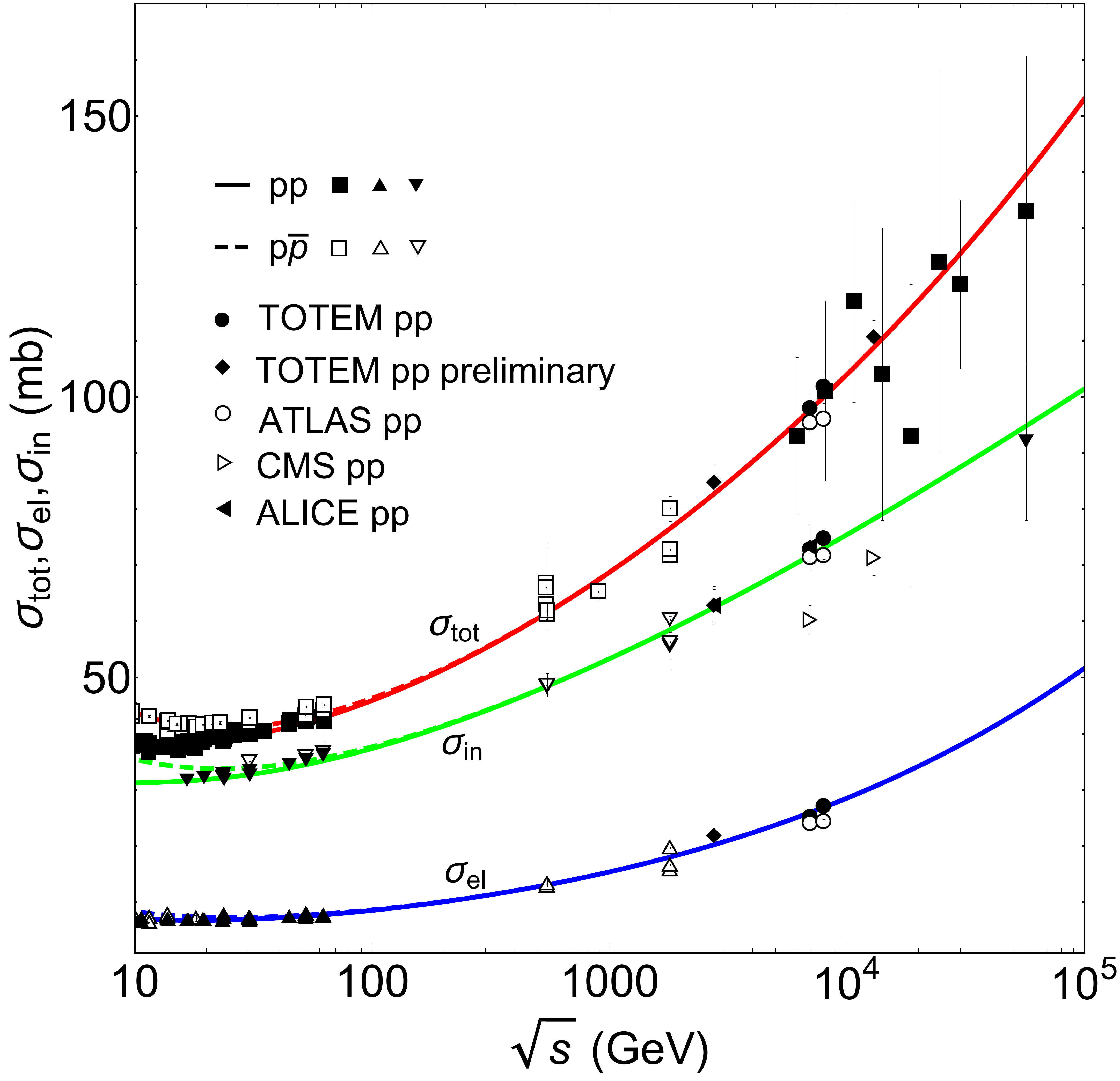}%
	}\hfil
	\subfloat[\label{fig:rho}]{%
	\includegraphics[scale=0.19]{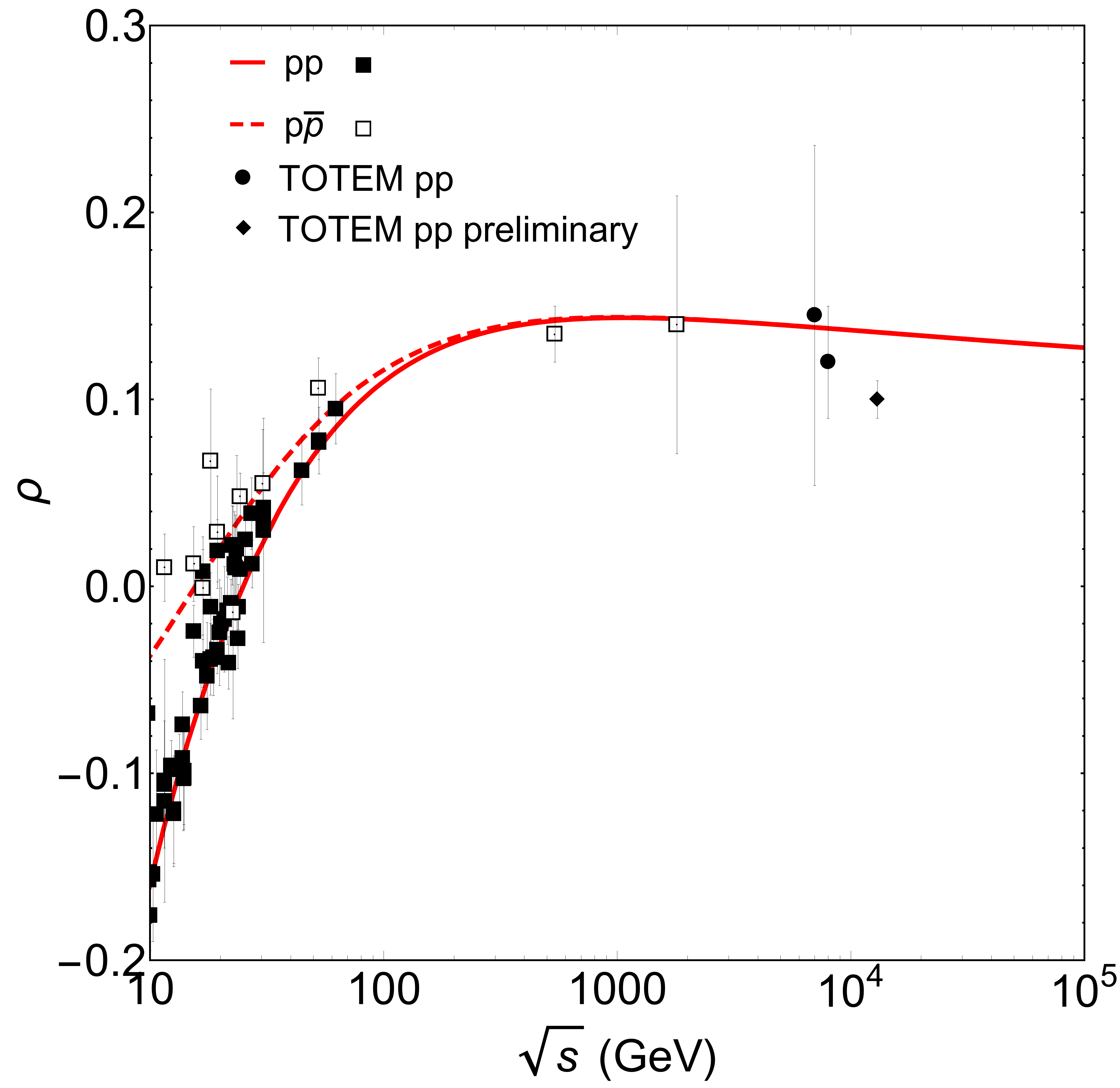}%
    }
	\caption{Fits to $pp$ and $\bar pp$ (a) total cross section data and (b) the ratio $\rho$ using Eqs.~(\ref{Eq:Amplitude}-\ref{Ptray}) without the odderon; calculated elastic and inelastic cross sections, Eqs.~(\ref{eq:el}-\ref{eq:inel}) are also shown.}
	\label{Fig:sigmarho}
\end{figure}

Fig.~\ref{Fig:sigmarho} shows the results of our fits to $pp$ and $\bar pp$ total cross section and the ratio $\rho(s,0)$ data.The data are from Refs.\cite{Csorgo,PDG,data,atlas7,atlas8,totem7,totem81,totem82,totem83,Auger}. The values of the fitted parameters and relevant values of $\chi^2/dof$ are presented in Table~\ref{tab:parameters}. 

\begin{table}[tbph!]
	\begin{center}
		\begin{tabular}{|c|c|c|c|}
			\hline 
			\multicolumn{2}{|c|}{Pomeron} &  \multicolumn{2}{|c|}{Reggeons} \\
			\hline 
			$a_P$ & $301\pm0.78$  & $a_f$ & $-16.4\pm0.061$ \\
			$b_P$ [GeV$^{-2}$] & $9.91\pm0.049$  & $b_f$ [GeV$^{-2}$] & $4.22\pm0.055$          \\
			$\delta_P$ & $0.0458\pm0.00011$ & $ - $ & $ - $ \\
			$\alpha_{1P}$ [GeV$^{-2}$]& $0.394\pm0.002$  & $a_{\omega}$ & $9.71\pm0.093$ \\
			$\alpha_{2P}$ [GeV$^{-1}$]  & $0.0148\pm0.00073$ & $b_{\omega}$ [GeV$^{-2}$] & $8 (fixed)$ \\
			$\varepsilon_P$ & $0$  & $ - $ & $-$         \\
			$s_{0P}$ [GeV$^2$] & $100$ &  $s_0$ [GeV$^2$] & $1$\\ \hline 
			\multicolumn{4}{|c|}{$\chi^2/dof=3.2$} \\
			\hline
		\end{tabular}
	\end{center}
	\caption{Fitted parameters to $pp$ and $p\bar p$ data on elastic differential cross section, total cross section and the parameter $\rho$.}
	\label{tab:parameters}
\end{table}

Elastic cross section $\sigma_{el}(s)$ is calculated by integration
\begin{equation}\label{eq:el}
\sigma_{el}(s)=\int_{t_{min}}^{t_{max}}\frac{d\sigma}{dt}(s,t)\, dt,
\end{equation}
whereupon
\begin{equation}\label{eq:inel}
\sigma_{in}(s)=\sigma_{tot}(s)-\sigma_{el}(s). 
\end{equation}
Formally, $t_{min}=-s/2$ and $t_{max}=t_{threshold}$, however since the integral is saturated basically by the first cone, we set $t_{max}=0$ and $t_{min}=-1$ GeV$^2$. The results are shown in Fig.~\ref{fig:sigma}.

\begin{figure}[H] 
	\centering
	\subfloat[\label{fig:elin}]{%
	\includegraphics[scale=0.19]{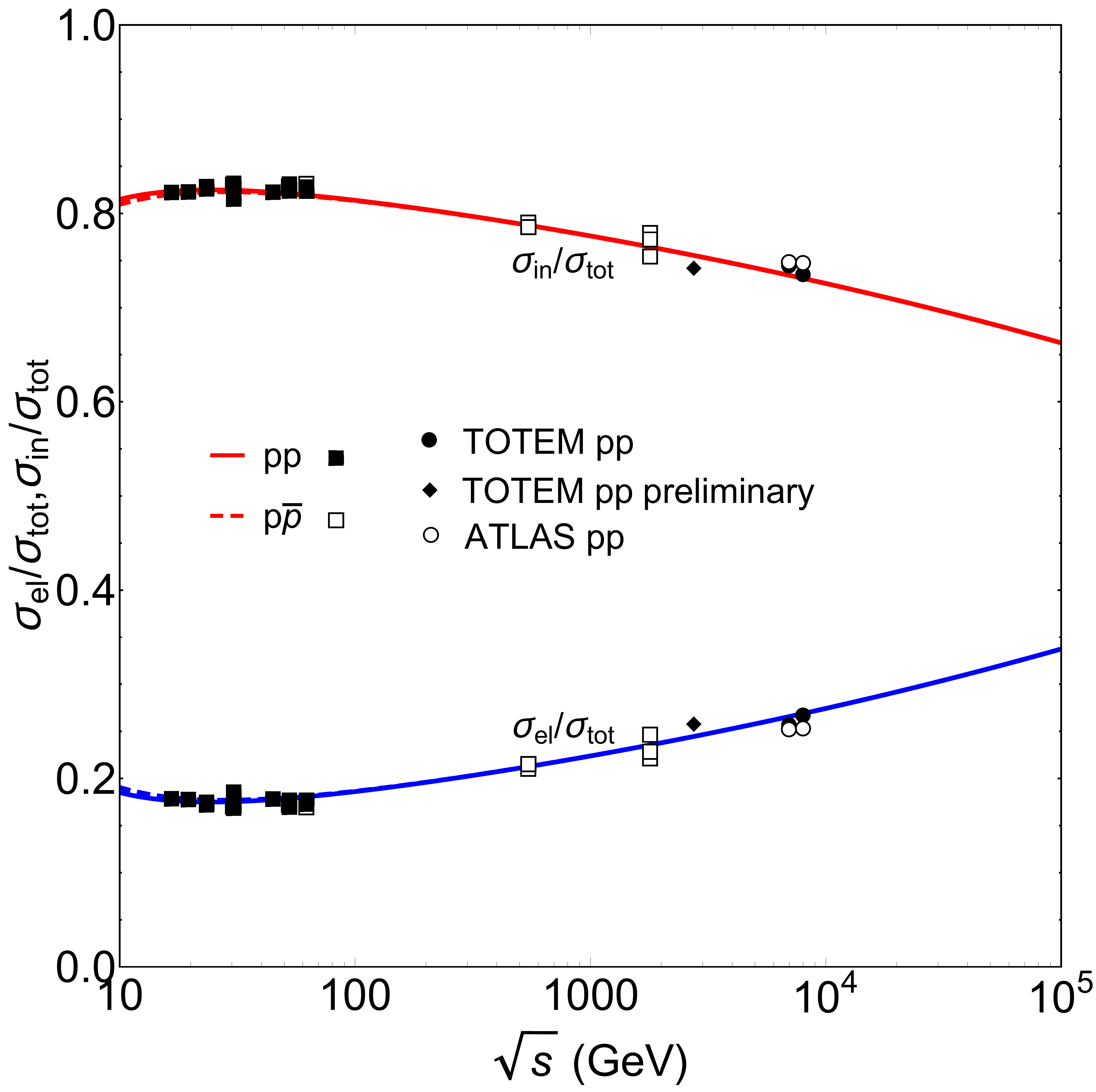}%
    }\hfil
	\subfloat[\label{fig:inel}]{%
	\includegraphics[scale=0.19]{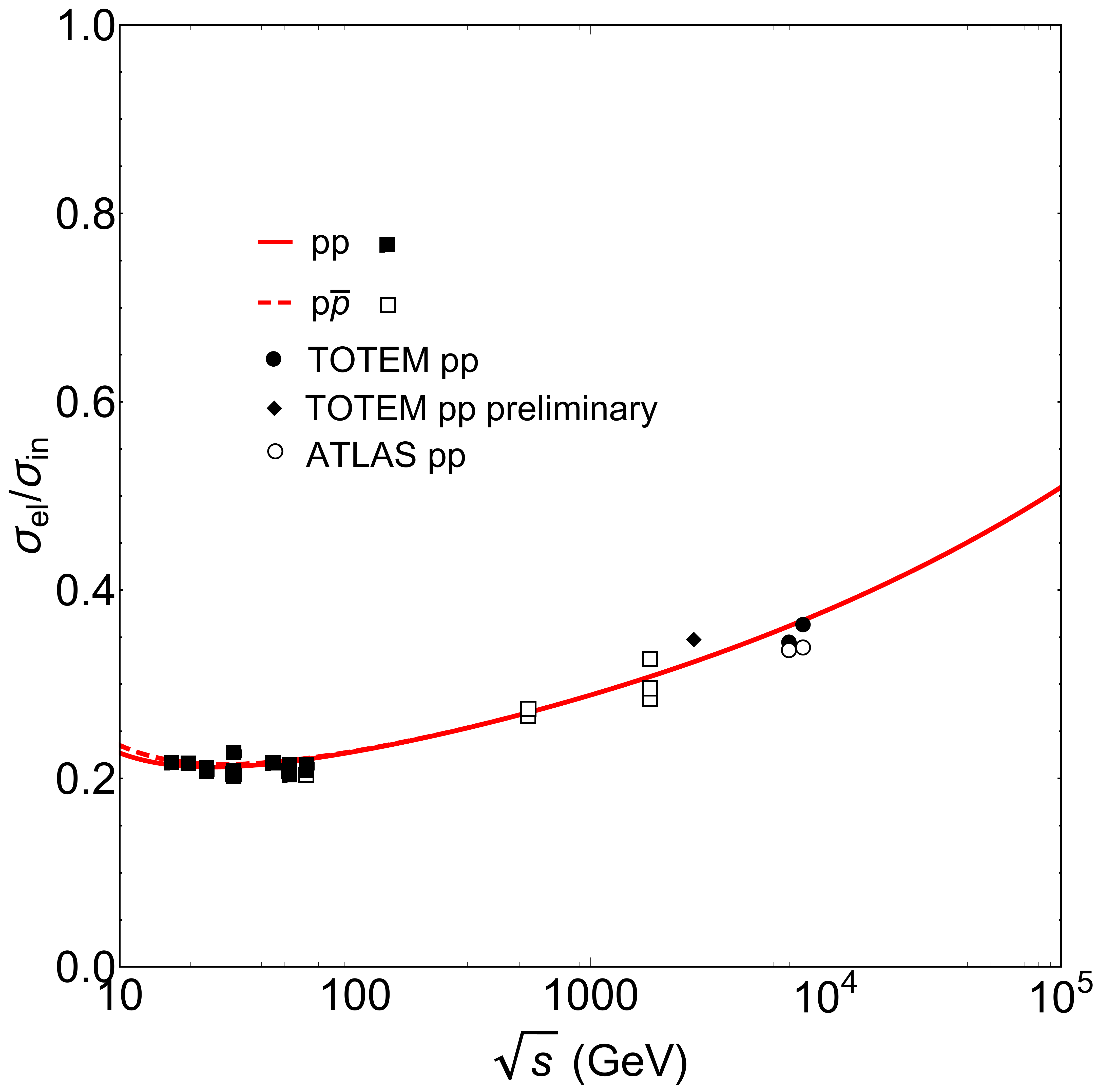}%
    }
	\caption{$\sigma_{el}/\sigma_{tot}$, $\sigma_{in}/\sigma_{tot}$ (a) and $\sigma_{el}/\sigma_{in}$ (b) ratios calculated from the fitted model, Eqs.~(\ref{Eq:Amplitude}-\ref{Ptray}), without the odderon.}
	\label{Fig:sratios}
\end{figure}

The calculated ratios of $\sigma_{el}/\sigma_{tot}$, $\sigma_{in}/\sigma_{tot}$ and $\sigma_{el}/\sigma_{in}$ are shown in Figs. \ref{Fig:sratios}. 
\begin{figure}[H] 
	\centering
	\subfloat[\label{fig:dsigmapp}]{%
	\includegraphics[scale=0.19]{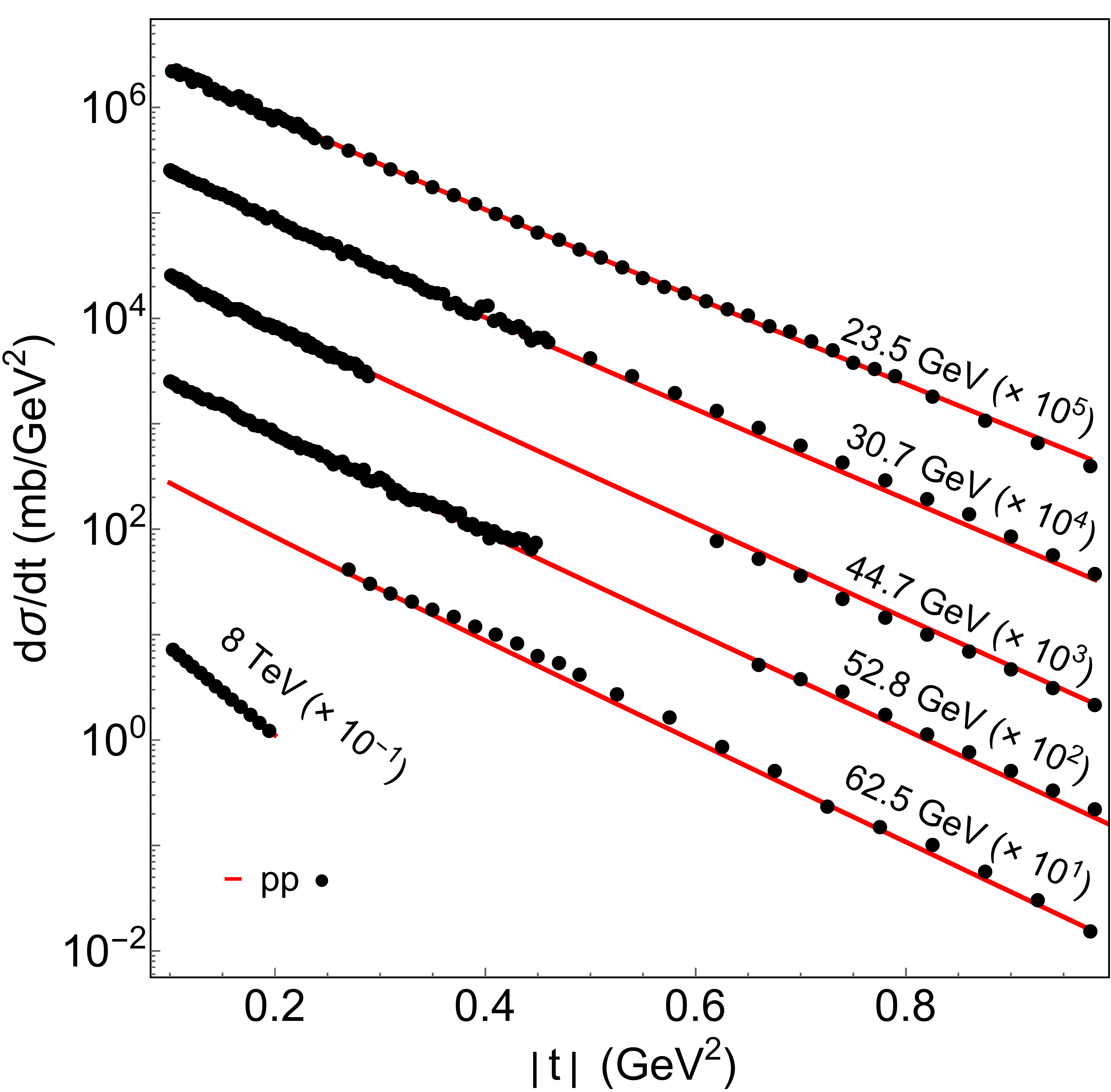}%
    }
	\hfil
	\subfloat[\label{fig:dsigmabarpp}]{%
	\includegraphics[scale=0.19]{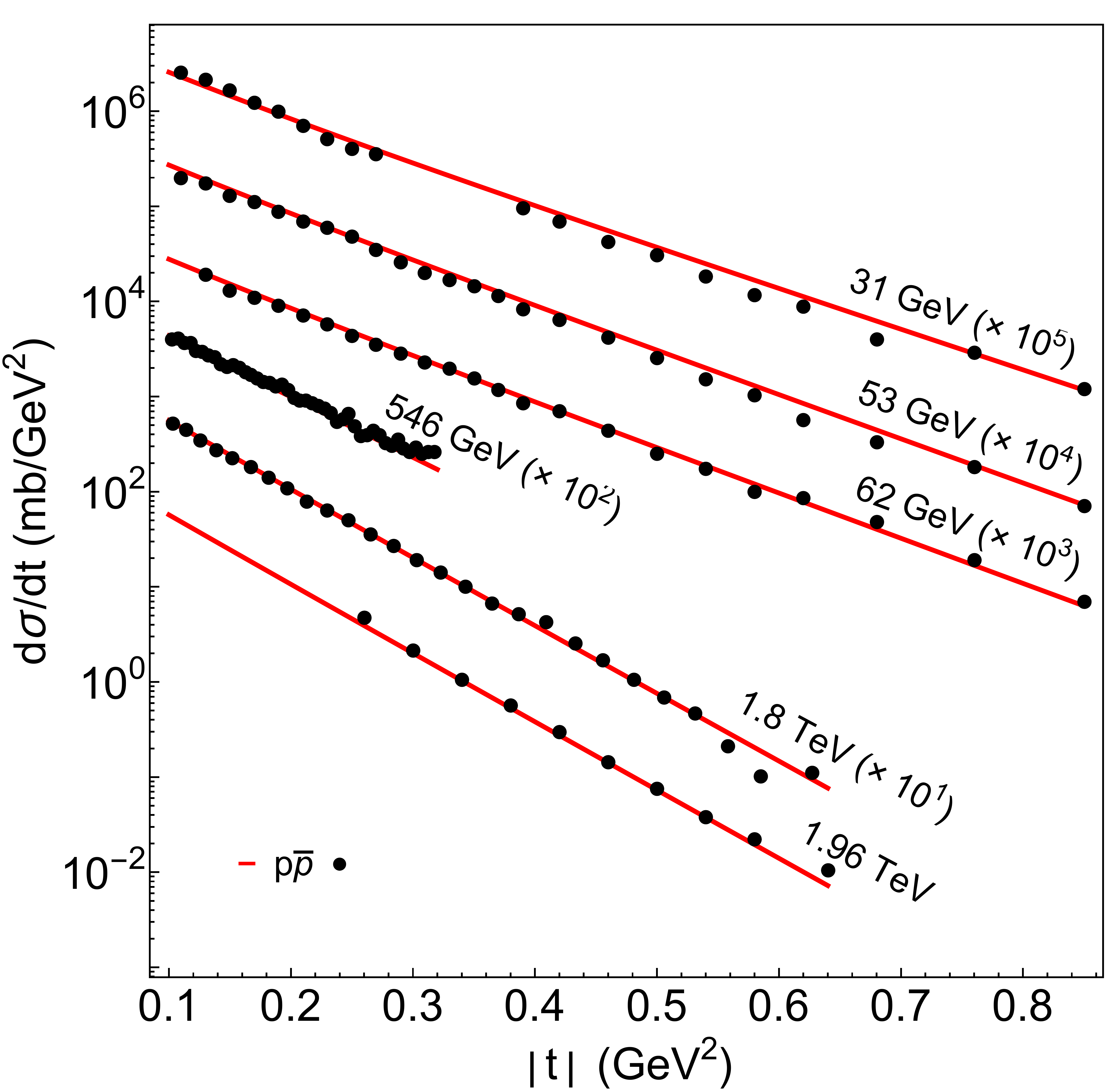}%
    }
	\caption{Results of our fit: (a) for $pp$ and (b) $\bar pp$ differential cross sections using Eqs.~(\ref{Reggeon1})-(\ref{Ptray}) without the odderon.}
	\label{Fig:dsigma}
\end{figure}

The above calculations are intended to provide the ground for our study of the slope, to be presented in the next Subsection. 

\subsection{Slope of the diffraction cone $B(s,t)$}\label{ssec:B}
With the model and its fitted parameters in hand, we now proceed to study the slope $B(s,t)$. Relevant formulas are presented in the Appendix. 

\begin{figure}[H] 
	\centering
	\subfloat[\label{fig:B}]{%
	\includegraphics[scale=0.19]{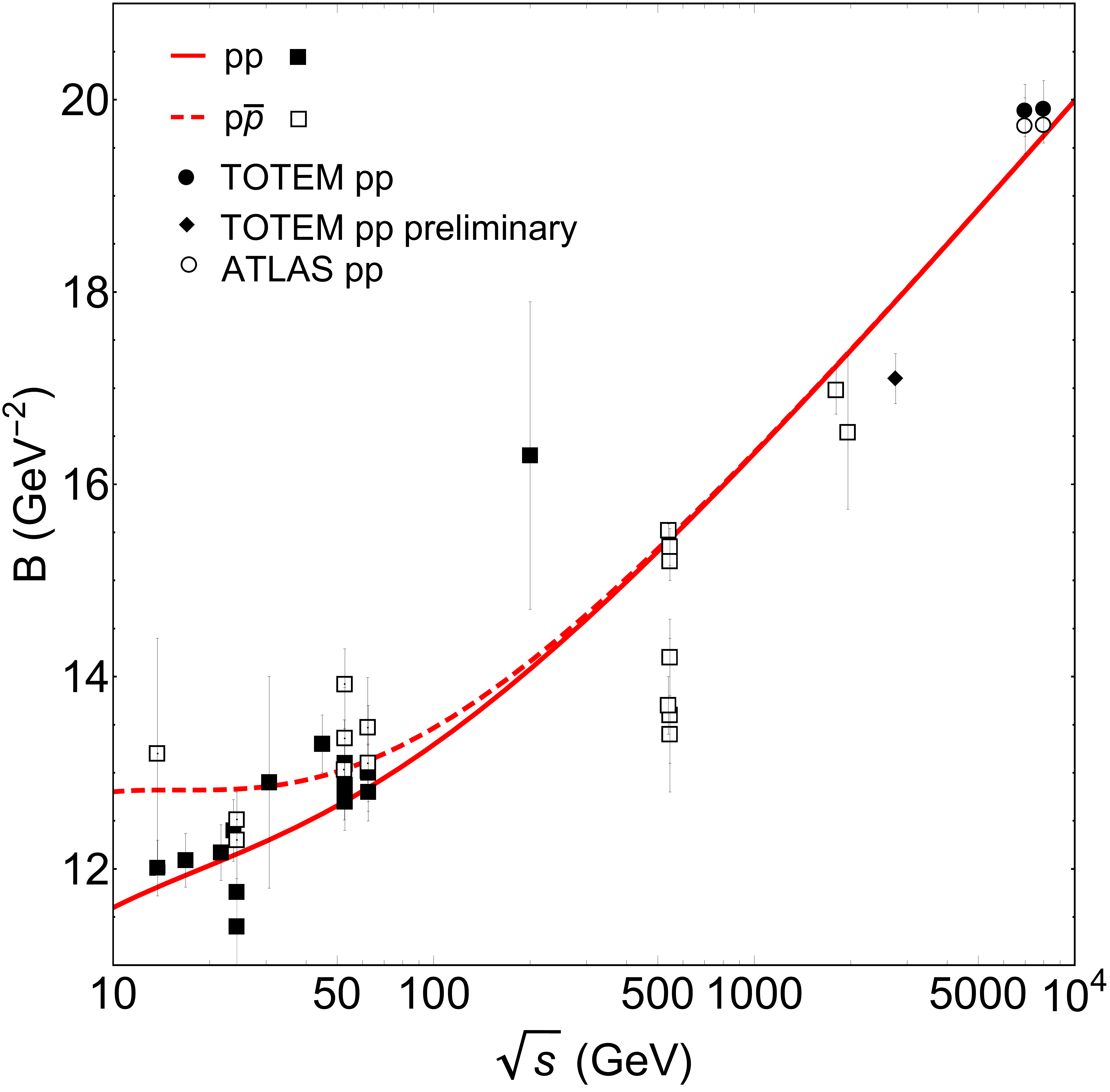}
    }\hfil
	\subfloat[\label{fig:Bsigmat}]{%
	\includegraphics[scale=0.19]{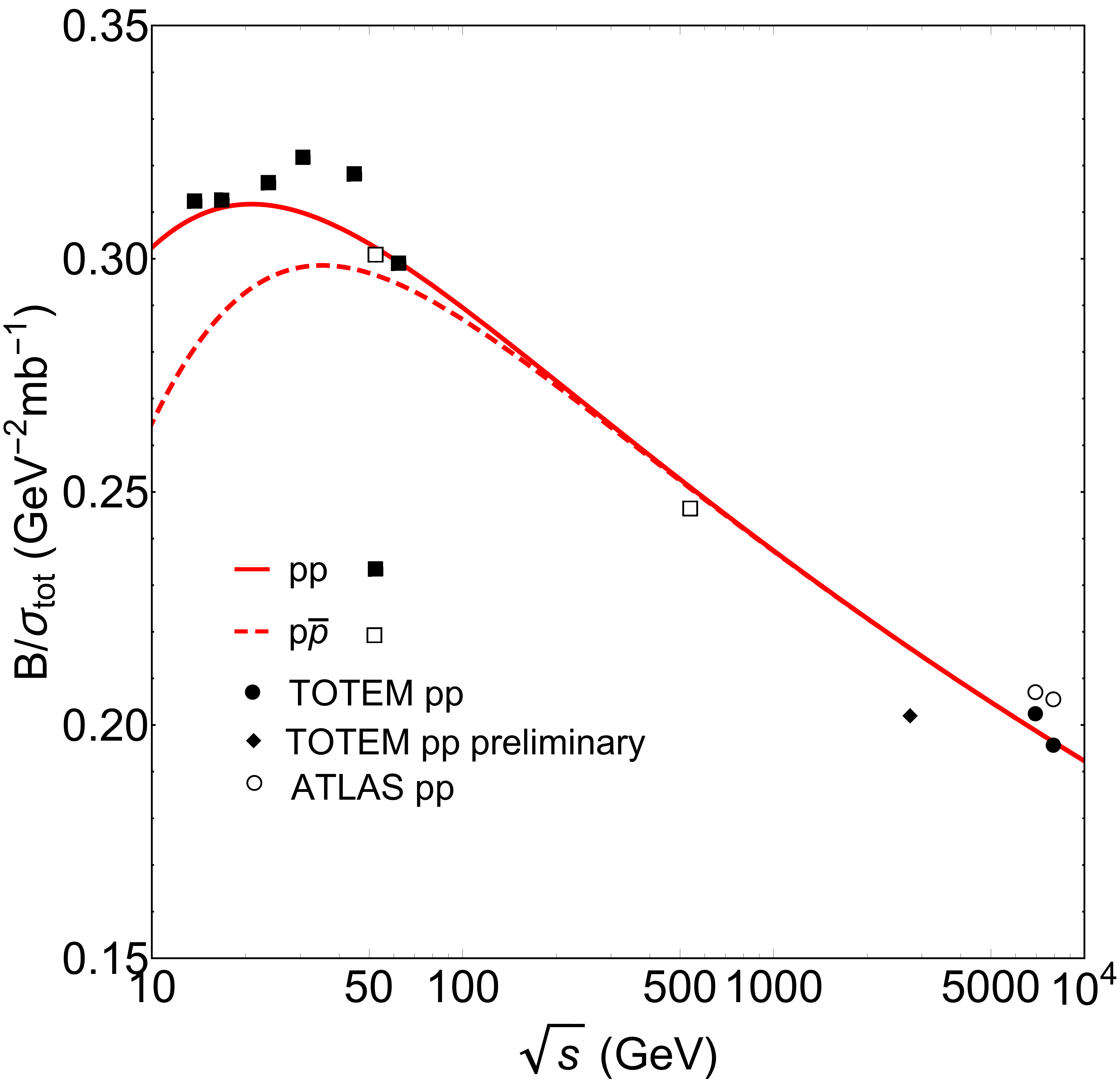}%
    }
	\caption{Calculated $pp$ and $\bar pp$ (a) elastic slope $B(s,0)$ and (b) $B(s,t=0)/\sigma_{tot}$ ratio from the fitted model Eqs.~(\ref{Eq:Amplitude})-(\ref{Ptray}) (without the odderon).}
	\label{Fig:B}
\end{figure}

\begin{figure}[H] 
	\centering
	\subfloat[\label{fig:Btpp}]{%
	\includegraphics[scale=0.19]{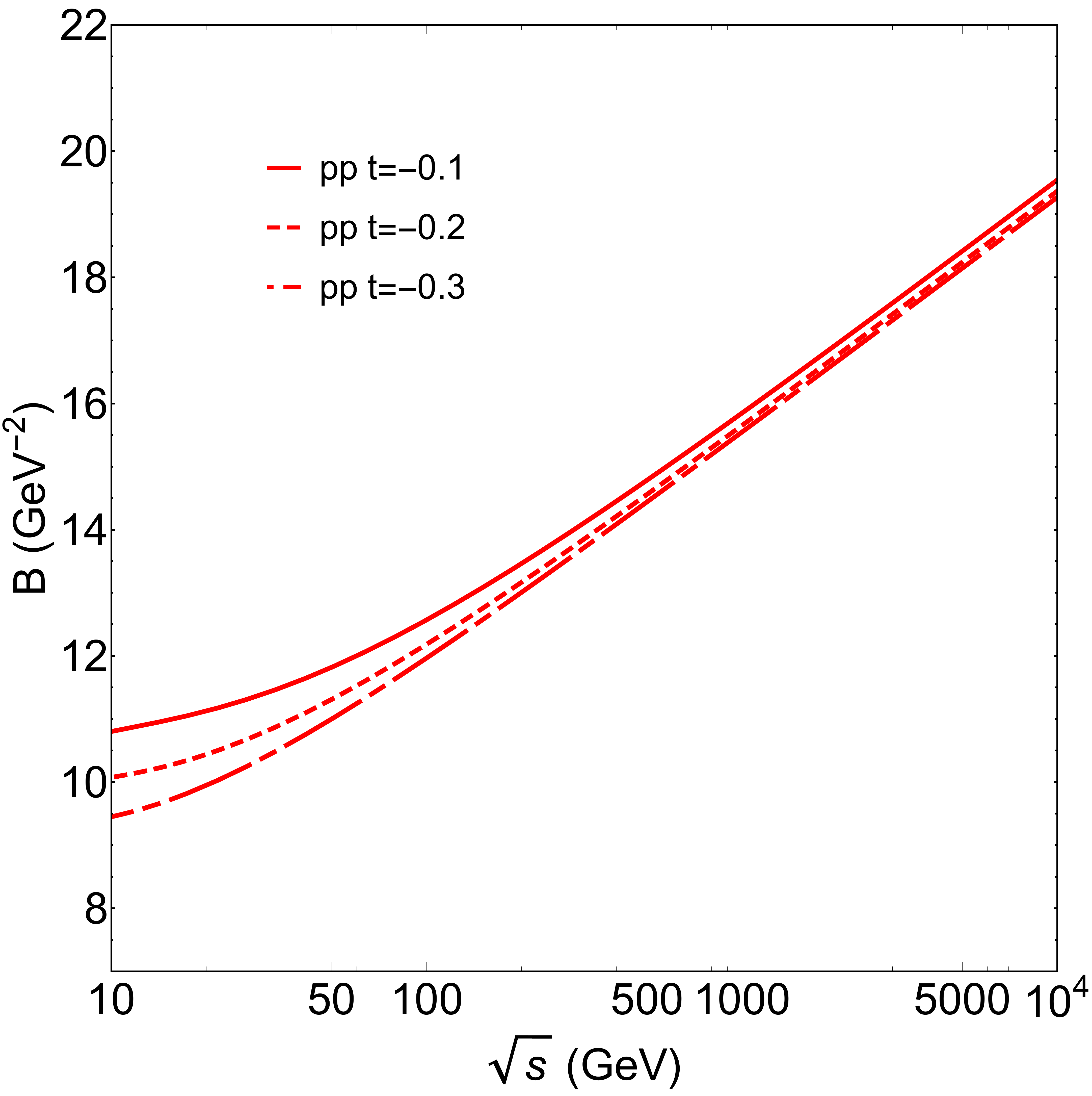}%
    }\hfil
	\subfloat[\label{fig:Btbarpp}]{%
	\includegraphics[scale=0.19]{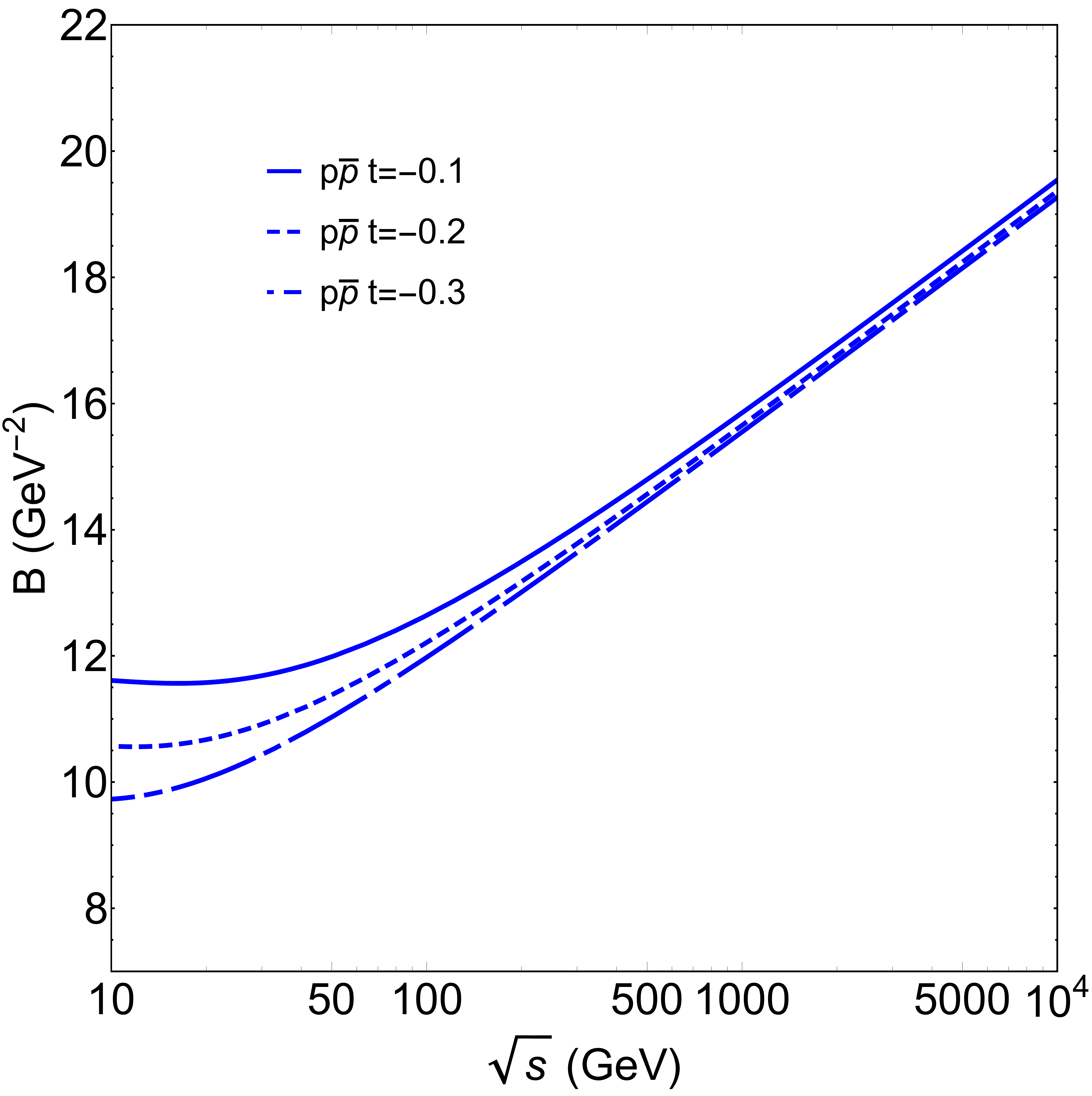}%
    }
	\caption{ (a) $pp$ and (b) $\bar pp$ elastic slope $B(s,t)$ Eq.~(\ref{Eq:slope}) calculated for several values of $t$ from  Eqs.~(\ref{Eq:Amplitude}-\ref{Ptray}) (without odderon).}
	\label{Fig:Bt}
\end{figure}

The elastic slope $B(s,0)$ and the ratio $B(s,0)/\sigma_{tot}(s)$ are shown in Fig.\ref{Fig:B}. Fig.\ref{Fig:Bt} shows the calculated $B(s,t)$ for various $t$ values. 

\begin{figure}[H] 	
	\centering
	\includegraphics[scale=0.19]{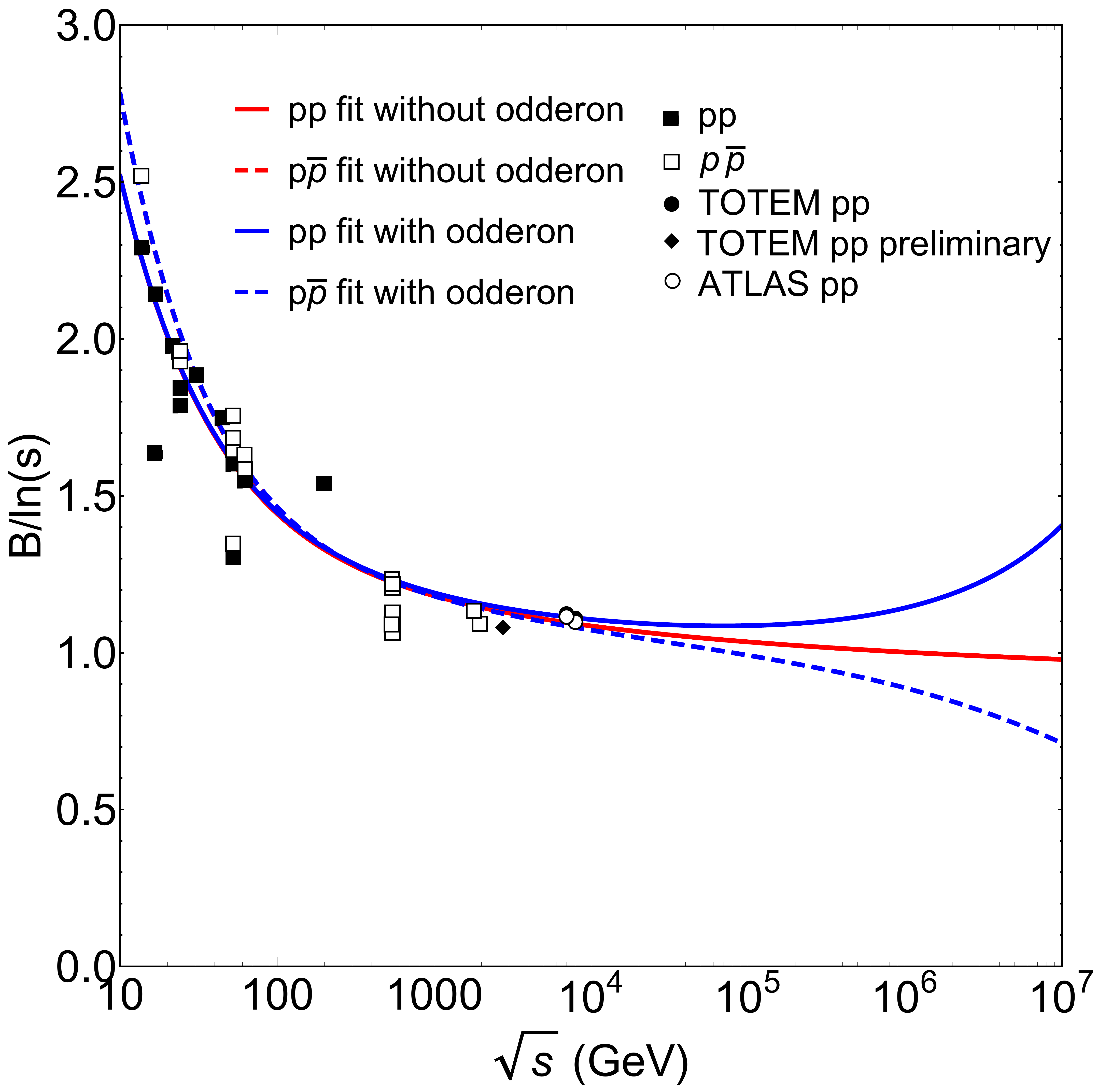}
	\caption{The ratio $B/ln(s)$ calculated from Eqs.~(\ref{Eq:Amplitude}-\ref{Eq:Otray}) with and without the odderon.}
	\label{Fig:Bo}
\end{figure}

To see better the effect of the odderon the deviation of $B(s)$ from its "canonical", logarithmic form, we show in Fig.\ref{Fig:Bo} its "normalized" shape, $B(s)/\ln(s)$. A similar approach was useful in studies \cite{totem83, Break} of the fine structure (in $t$) of the diffraction cone.

More detail from the slope given by Eq.~(\ref{Eq:slope}) with the norm Eq.~(\ref{norm}) and the amplitude Eq.~(\ref{Eq:Amplitude}) can be find in the Appendix.   

The main conclusion from this section is that the dipole pomeron at the "Born" level, fitting data on elastic, inelastic and total cross section, does not reproduce the irregular behavior of the forward slope observed at the LHC. Account for unitarity is indispensable. Quite interestingly, the inclusion of the odderon reproduces the $ln^2(s)$ behavior of the elastic slope beyond the LHC energy region.

\section{Unitarity}\label{Sec:Unitarity}

We use a unitarization procedure by which the unitarized amplitude is 
\begin{equation}\label{Eq:U}
T(\rho, s)=\frac{u(\rho,s)}{1-iu(\rho,s)}.
\end{equation}
Here the input $u(\rho,s)$, analogue of the eikonal, is the Fourier-Bessel transform of the Regge-pole amplitude, the DP Eq. (\ref{GP}) in our case. The advantage of this unitarization procedure, called (somewhat misleadingly) "u-matrix", is its simplicity (rational functions instead of exponentials in eikolalization \cite{Martin}), see \cite{TT} and references therein. 

For $t=0$ calculations can be done analytically in terms of a series in $1/L$. After a Fourier-Bessel transform over the DP amplitude Eq.~(\ref{GP}), Sec. \ref{Sec:DP} we get the $u(\rho,s)$. By keeping terms and introducing the variable $x=\rho^2/4\alpha'_PL$ (where $\alpha'_P$ is the derivative of the pomeron trajectory), one gets \cite{PEPAN} up to $O(1/L)$
\begin{equation}
u(x,s)=ige^{-\chi}(1+\chi x),
\end{equation}
where
\begin{equation}
g=\frac{\sigma_0\lambda}{4\pi\alpha'_P},\ \ \chi=\Bigl(1+i\pi\lambda/2\Bigr)/{\lambda L},\ \ \lambda=(1-\epsilon_P)/b_P, \ \ \sigma_0=\frac{4\pi}{s_{0P}}G'(0),
\end{equation}
whereupon the profile function assumes the form
\begin{equation}\label{Eq:proffun}
\frac{u}{1-iu}=\frac{ige^{-\chi}}{1+ge^{-\chi}}\Bigl(1+\chi\frac{x}{1+ge^{-\chi}}\Bigr).
\end{equation}
by an inverse Fourier-Bessel transform over Eq.~(\ref{Eq:proffun}) we obtain for forward measurables, in the $O(1/L)$ approximation:

\begin{equation}\label{Eq:tot}
\sigma_{tot}=\frac{4\pi\alpha'_P}{\lambda}\ln(1+g)(1+\lambda L).
\end{equation}

\begin{equation}\label{Eq:in}
\sigma_{in}=\frac{4\pi\alpha'_P}{\lambda}\frac{g}{1+g}(1+\lambda L).
\end{equation}

\begin{equation}\label{Eq:el}
\sigma_{el}=\frac{4\pi\alpha'_P}{\lambda}(\ln(1+g)-\frac{g}{1+g})(1+\lambda L).
\end{equation}

One can see from Eqs.~(\ref{Eq:tot})-(\ref{Eq:el}) than, in the leading $O(1/L)$ approximation, the energy dependence of the cross sections is not affected by unitarization if $g$ is constant, that was typical for the ISR era with geometrical scaling (GS). GS is violated beyond the ISR, requiring energy dependence in $g\rightarrow g(s),$ to be discussed in what follows. 

Furthermore, in the $O(1/L)$ approximation
\begin{equation}
A(s,0)=A_B(s,0)\frac{ln(1+g)}{g},
\end{equation}
The ratio of the real and imaginary part of the scattering amplitude:
\begin{equation}
\frac{ReA(s,0)}{ImA(s,0)}=\frac{\pi\lambda}{2(1+\lambda L)}=\frac{ReA_B(s,0)}{ImA_B(s,0)},
\end{equation}
where
\begin{equation}
A_B(s,0)=i\frac{\sigma_0s}{4\pi}\Bigl[1+\lambda\Bigl(L-i\pi/2\Bigr)\Bigr].
\end{equation}
For the slope one gets
\begin{equation}\label{Eq:UniB}
B(s,0)=\frac{2\alpha'_P}{\lambda}\frac{\sum}{\ln(1+g)}(1+\lambda L),
\end{equation}
where
\begin{equation}\label{Eq:SumB}
\sum=\int_{0}^{\infty}	\frac{ge^{-x}xdx}{1+ge^{-x}}.
\end{equation}

Note that for small $x$ this integral may be replaced by the sum

\begin{equation}
\sum=g\sum_{n=0}^{\infty} (-g)^{n}(n+1)^{-2}.
\end{equation}

Hence the relation

\begin{equation}
B(s,0)=k\alpha'_P\sigma_{tot},
\end{equation}

where

\begin{equation}
k=\frac{\sum}{2\pi\alpha'_P\ln^{2}(1+g)}.
\end{equation}

For linear trajectories, it may be rewritten as 

\begin{equation}
B(s,0)=\frac{1}{2\pi}\frac{\sum}{\ln^{2}(1+g)}\sigma_{tot}.
\end{equation}

In the "Born approximation" we have the McDowell-Martin limit:
\begin{equation}
B(s,0)=\frac{\sigma_{tot}^2}{4\pi\sigma_{el}},
\end{equation}
 
Note that in the above derivations, smallness of $g$ was never required. 

The parameter $g$ may be found from the ratio
\begin{equation}\label{Eq:ge}
\frac{\sigma_{el}}{\sigma_{tot}}=1-\frac{g}{(1+g)\ln(1+g)}.
\end{equation}
where $g$ is constant in the case of unit DP intercept, sharing the property of geometrical scaling (GS), typical of the ISR energy region, with 
\begin{equation}
\sigma_t\sim\sigma_{el}\sim\sigma_{in}\sim B(s)\sim\ln(s).
\end{equation}

 Beyond the ISR, GS is broken and the DP becomes supercritical.

\begin{figure}[H]\label{Fig:g}
	\centering	
	\subfloat[\label{fig:g}]{%
	\includegraphics[scale=0.19]{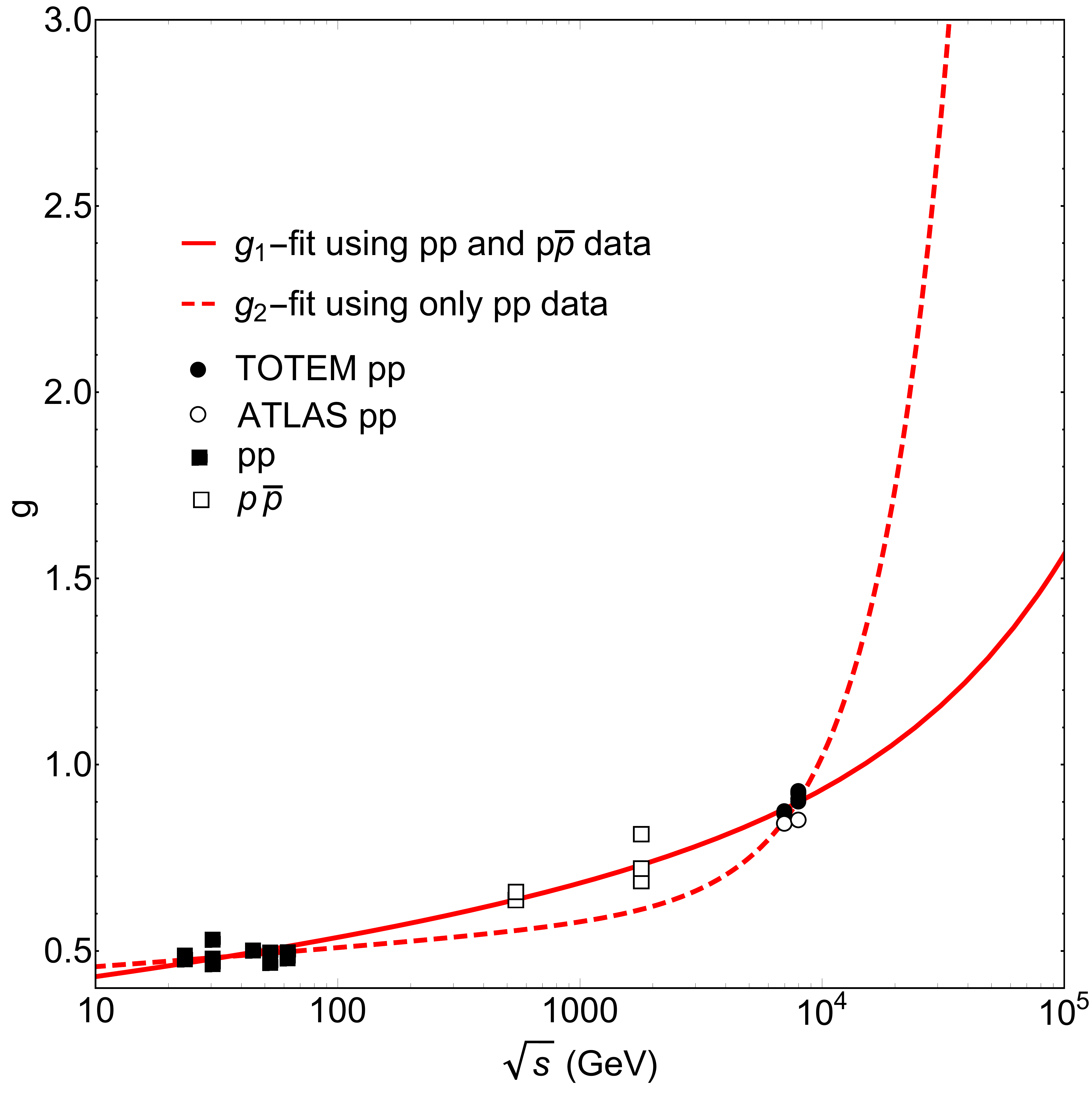}%
    }\hfill
	\subfloat[\label{fig:SU}]{%
	\includegraphics[scale=0.19]{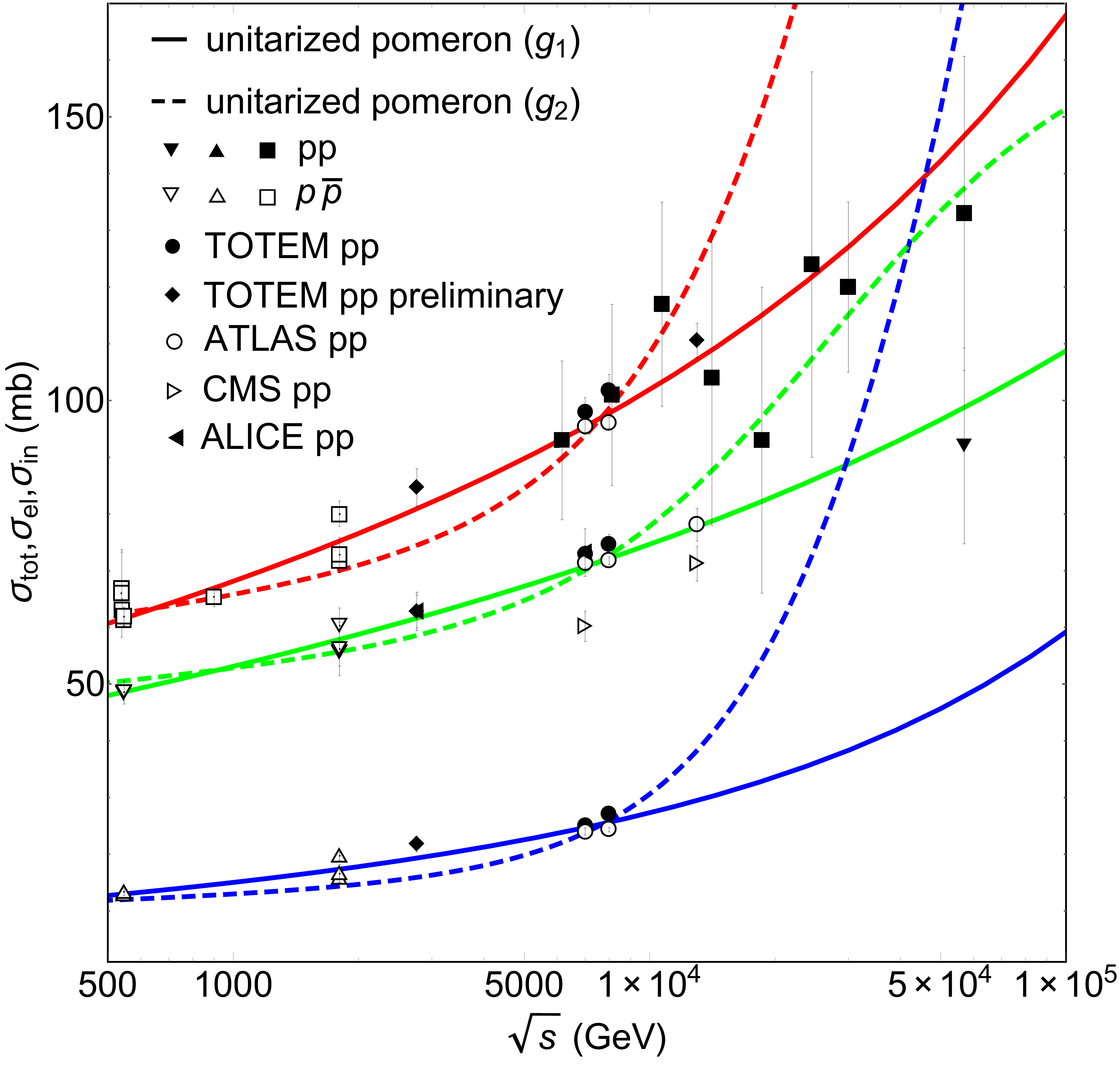}%
    }
	\caption{Fitted values of $g$, Eq.~(\ref{Eq:ge}), Eq.~(\ref{Eq:gpar}) and the resulting cross sections Eqs.~(\ref{Eq:tot}-\ref{Eq:tot}).}
\end{figure}

From Eq.~(\ref{Eq:ge}) we calculate $g$ using the $pp$ and $\bar pp$ data. By parametrizing
\begin{equation}\label{Eq:gpar}
g(s)=g_{01}(s/s_{01})^{\epsilon_1}+g_{02}(s/s_{02})^{\epsilon_2} 
\end{equation}
we fitted $g(s)$ to the data with two options: 1) include both $pp$ and $\bar pp$ "$g$ data", denoted the fitted function $g_1(s)$; 2) include only $pp$ data, denoting the fitted function $g_2(s)$. The results are shown in Fig.~\ref{fig:g} with the values of the fitted parameters presented in Tab.~\ref{tab:fitParam1}.

\begin{figure}[H] \label{FigBU}
	\centering
	\subfloat[\label{fig:uniB}]{%
	\includegraphics[scale=0.19]{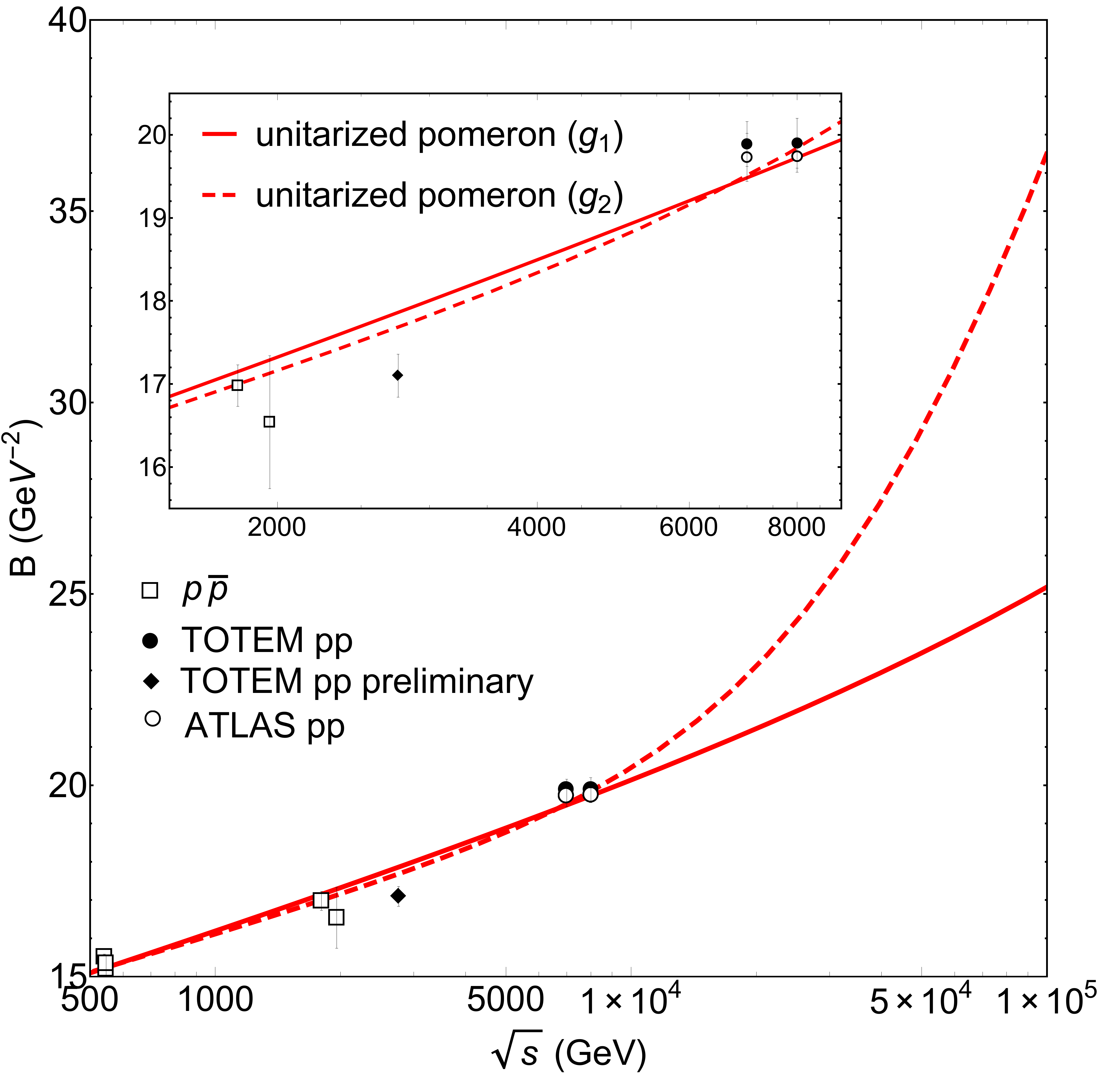}%
	}\hfill
	\subfloat[\label{fig:test}]{%
	\includegraphics[scale=0.19]{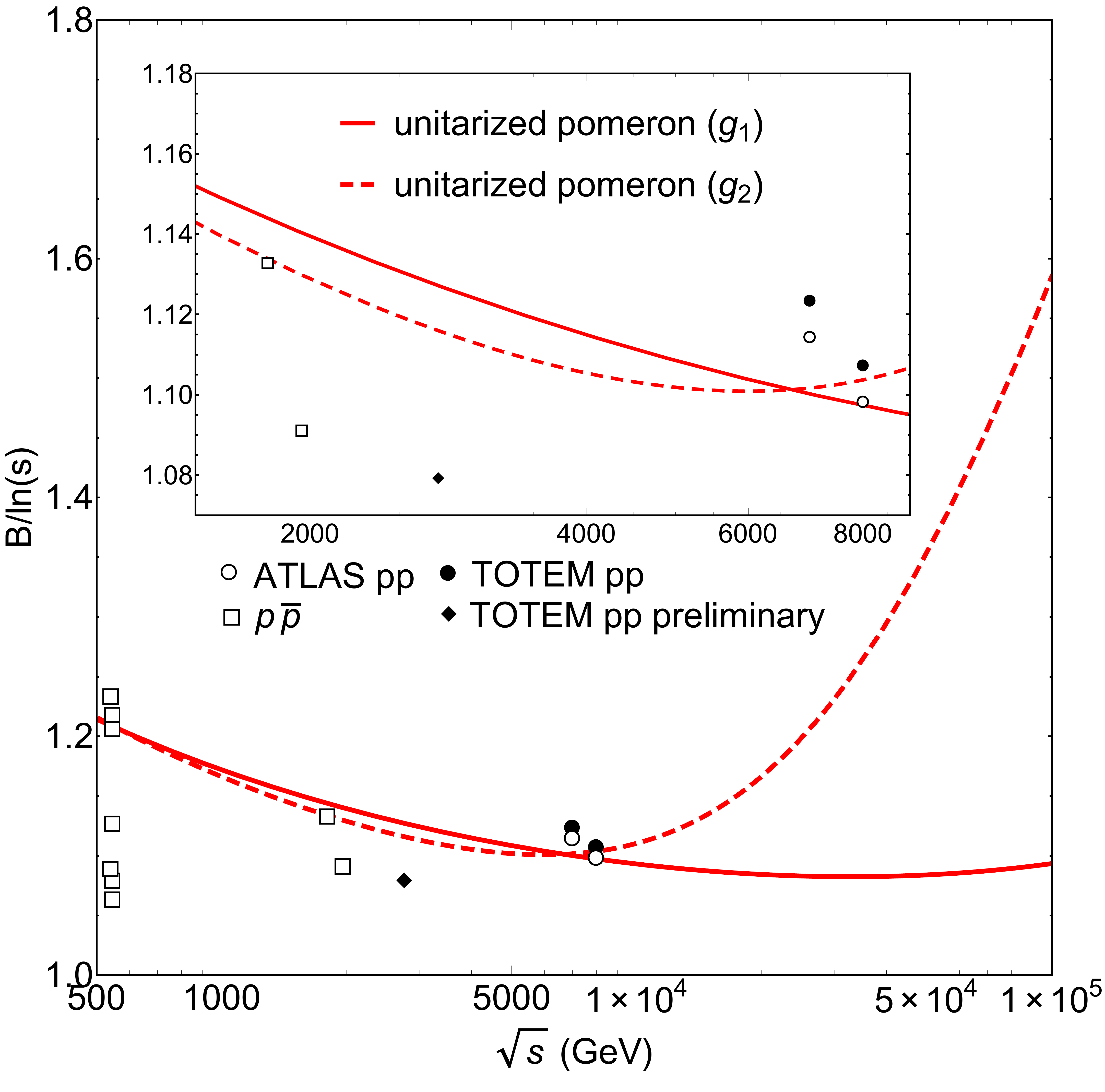}%
	}   
	\caption{(a) Elastic slope calculated from Eq.~(\ref{Eq:UniB}) and (b) the "test" ratio $B/ln(s)$.}
\end{figure}

 Using the unitarized formulas with the obtained $g_1(s)$ and $g_2(s)$ we calculate the elastic slope and the total, elastic and inelastic cross sections by treating $\alpha'_P$ and $\lambda$ as free parameters fitted to the data. The values of the obtained parameters are shown in Tab.~\ref{tab:fitParam1}. The unitarized cross sections are shown in Fig.\ref{fig:SU}; the unitarized version of the elastic slope is shown in Fig.~\ref{fig:uniB} and its "normalized" form in Fig.~\ref{fig:test}.

\begin{table}[tbph!]
	\begin{center}
        \subfloat[$g(s)$ fitted to $pp$ and $p\bar p$ data \label{tab:uni2}]{%
\begin{tabular}{||c|c|c|c|c|c||c|c||c|c||}
	\hline
	\multicolumn{6}{|c|}{$g_1(s)$} & \multicolumn{2}{|c|}{$B(s)$} & \multicolumn{2}{|c|}{$\sigma_{tot,el,in}(s)$} \\\hline
	$g_{01}$&$0.348$ &$\epsilon_1$&$0.0457$ & $s_{01}$ & $1$ (fixed)& $\alpha'_P$ & $0.808$ & $\alpha'_P$ & $1.05$ \\
	\hline
	$g_{02}$&$0.00135$ &$\epsilon_2$&$0.328$ &$s_{02}$& $100$ (fixed)& $\lambda$ & $0.0737$ & $\lambda$  & $0.0613$ \\
	\hline
\end{tabular}%
        }\qquad
    \subfloat[$g(s)$ fitted only to $pp$ data \label{tab:uni1}]{%
    	\begin{tabular}{||c|c|c|c|c|c||c|c||c|c||}
    		\hline
    		\multicolumn{6}{|c|}{$g_2(s)$} & \multicolumn{2}{|c|}{$B(s)$} & \multicolumn{2}{|c|}{$\sigma_{tot,el,in}(s)$} \\\hline
    		$g_{01}$&$0.412$ &$\epsilon_1$&$0.0228$ & $s_{01}$ & $1$ (fixed)& $\alpha'_P$ & $0.778$ & $\alpha'_P$ & $0.434$ \\
    		\hline
    		$g_{02}$&$0.0000153$ &$\epsilon_2$&$0.735$ &$s_{02}$& $100$ (fixed)& $\lambda$ & $0.0682$ & $\lambda$  & $0.0170$    \\
    		\hline
    	\end{tabular}%
    }
	\end{center}
	\caption{Values of the fitted parameters in $g(s)$, the slope as well as total, elastic and inelastic cross sections.}
	\label{tab:fitParam1}
\end{table}


One can see  from Fig.~\ref{fig:uniB} and Fig.~\ref{fig:test} that the unitarized pomeron with $g_1(s)$ reproduces the $ln^2s$ rise for the slope starting from $\sqrt{s}\approx 13$ TeV; with $g_2(s)$ it produces a $ln^2s$ rising slope starting already from $\sqrt{s}\approx 5$ TeV, in agreement with the recent experimental results \cite{Oster,Csorgo,Mario,Giani}. 

\section{"Asymptopia"}\label{Sec:Asympt}
\subsection{Asymptotic universality}

In this Section we use the obtained results to predict future trends. Our first conclusion is the existence of two regimes: 1) the first is the "low-energy" ISR-FNAL region, shows modest (logarithmic) rise of the cross section with constant ratios $\sigma_{el}/\sigma_{tot}$ and $\sigma_{tot}/B$ (geometrical scaling, GS); 2) the next is asymptotic, with Froissart saturation and $\sigma_{el}/\sigma_{tot}\rightarrow 1$. The transition between the two and onset of the asymptotic regime is quantized by the "running constant" $g(s)$, parametrized in the previous section.

Having fixed $g(s)$ and using Eq.(\ref{Eq:tot}) that the rise of $\sigma_{tot}$ is a combined effect coming from two factors: increasing intensity of the interaction ${\cal L}_1=ln[1+g(s)]$  and increasing interaction radius ${\cal L}_2=1+\lambda L.$ Their product results in Froissart saturation,
\begin{equation}\label{Eq:Froissart}
\sigma_{tot}\sim c\ln^2 s.
\end{equation} 

In the GS (ISR) region, the $\ln^2(s)$ component makes merely $10\%$ of the total cross section, rescattering corrections becoming important only as $g_1(15 TeV)=1$ or $g_2(10 TeV)= 1 $. The asymptotic $\ln^2(s)$ regime sets on beyond $g>1$. For example, $g$ reaches $10$ at very high energy, $\sqrt{s}=6000$ TeV in case of $g_1$ or $\sqrt{s}=85$ TeV in case of $g_2$. 

The asymptotic behavior of the slope may be predicted from Eq. (\ref{Eq:SumB}). Expanding $g(s)$ at high energies, we rewrite $\sum$ as 
\begin{equation}
\sum(g)=\frac{1}{2}\ln^{2}g+\frac{\pi^{2}}{6}+\sum_{n=1}^{\infty}(-g)^{-n} n^{-2}.
\end{equation}
One sees that as $g>1$, the slope $B(s,0)$ increases as $L^2.$

The ratio $\sigma_{el}/\sigma_{tot}$ will cross the "critical" value $1/2$ at $\sqrt{s}=10^3$ TeV in case of $g_1$ or at $\sqrt{s}=40$ TeV in case of the $g_2$, whereupon it will tend to its asymptotic limit $\sigma_{el}/\sigma_{tot}\rightarrow 1,$ typical of the $u-$matrix approach \cite{TT}.

The ratio $\sigma_{tot}/B$ is also increasing tending to its asymptotic limit
\begin{equation}
\sigma_{tot}/B=4\pi.
\end{equation}

Note that our formulas show explicitly the transition from the logarithmic rise of the cross sections and of $B(s)$ to the Froissart regime with $\sigma_{tot}(s)\sim B(s) \sim \ln^2(s).$

Another interesting feature of the model is the universality of cross sections manifest at super-high energies. To show it, we rewrite Eq.~(\ref{Eq:tot}) as
\begin{equation}\label{Eq:asympt}
\sigma_{tot}=4\pi\alpha'\ln[1+g_{01}(s/s_{01})^{\epsilon_1}+g_{02}(s/s_{02})^{\epsilon_2}](1/\lambda+L).
\end{equation}

When $g(s)=g_{01}(s/s_{01})^{\epsilon_1}+g_{02}(s/s_{02})^{\epsilon_2}\gg 1$ and $L\gg 1$ (the first condition being stronger), Eq.~(\ref{Eq:asympt}) may be rewritten approximately as 
\begin{equation}
\sigma_{tot}\approx c\ln^2 s,\ \  c=4\pi\alpha'_P\delta_P.
\end{equation} 

The value of the coefficient in front of $\ln^2 s$ saturating the Froissart bound is a subject of debates in the literature. It may be related to the pion mass as $c=\pi/m^2_{\pi}\approx60$ mb, that is much bigger than the value obtained from fits to the data. 

Eq. (\ref{Eq:Froissart}) relates the coefficient $c$ with the slope of the pomeron trajectory $\alpha'_P$ (string tension) and the parameter $\delta_P$. Setting the fitted values of $\alpha'_P=0.394$ GeV$^{-2}$ and $\delta_P=0.0458$ we get $c=0.088$ mb, close to its fitted value. 

Asymptotically universal are also $\sigma_{el},\ \ \sigma_{in}$ and $B(s).$ 

 \subsection{Quark counting rules}
Eq.~(\ref{Eq:asympt}) makes it possible to identify the component of the total cross section responsible for the relation between cross sections of various reactions by the so-called quark counting rules.

The relations between total cross sections following from the quark counting rules are known to be valid approximately to about $10\%$, and they change with energy. In our opinion, the relations make sense only after proper separation of various components of the cross section. Our model makes possible their separation and identification: it is the first, constant term in ${\cal L}_2$ as well as $g_0$ in ${\cal L}_1$ who count the quarks, while the logarithmically rising component is related to the universal gluonic component. 
   
While at "low" energies the quark component is hidden by the exchange of secondary trajectories, at high energies it is masked by the logarithmically increasing gluonic component. There is however a "window", around the FNAL energy region, where the contribution from secondary reggeons is about $20\%$ and $L\ll 1/\lambda$. For example, at $\sqrt{s}=20$~GeV $L=\ln 400\approx6$, less than $1/\lambda\approx20$.

\subsection{Predictions}
In Tab.~\ref{Table.3} (using $g_1(s)$) and Tab.~\ref{Table.4} (using $g_2(s)$) we present our predictions for the forward measurables at future energies.

\begin{table}[H]
	\centering
	\subfloat{%
		\begin{tabular}{|c||c|c|c|c|c|c|c|c|c|}\hline
			$\sqrt{s}, TeV$&0.9 &2.76 &3 &4 &13 &14 &80 &90 &100 \\\hline \hline
			$g$	&0.67& 0.77&0.78&0.81&0.98&0.99& 1.46& 1.51& 1.56 \\\hline
			$\sigma_{tot} (mb)$	&66.9&80.9&82.0& 86.2& 107&108&158&163&167 \\\hline
			$\sigma{el} (mb)$	&14.6&19.3&19.7& 21.2& 29.4& 30.0&54.1&56.5& 58.9 \\\hline
			$\sigma{in} (mb)$	&52.3&61.6& 62.3&65.0&77.7&78.6&104&106&108 \\\hline
			$\sigma_{el}/\sigma_{tot}$	&0.218&0.238&0.240&0.246&0.274&0.276&0.341&0.313&0.351 \\\hline
			$B (GeV^{-2})$	&16.0&17.9&18.1&18.6&20.8&20.9&24.2&24.6&24.9 \\\hline
		\end{tabular}%
	}
	\caption{Predictions using the unitarization procedure with $g_1(s)$.}
	\label{Table.3}
\end{table}

\begin{table}[H]
	\centering
	\subfloat{%
		\begin{tabular}{|c||c|c|c|c|c|c|c|c|c|}\hline
			$\sqrt{s}, TeV$&0.9 &2.76 &3 &4 &13 &14 &80 &90 &100 \\\hline \hline
			$g$	&0.574& 0.65&0.66&0.70&1.2&1.3& 9& 11& 12 \\\hline
			$\sigma_{tot} (mb)$	&65.2&74.4&75.6& 80.0& 123&128&376&401&424 \\\hline
			$\sigma{el} (mb)$	&12.7&15.9&16.3&18.0&38.2& 41.0&229&252&272 \\\hline
			$\sigma{in} (mb)$	&52.4&58.5&59.3&62.0&85.0&87.3&146&149&151 \\\hline
			$\sigma_{el}/\sigma_{tot}$	&0.196&0.213&0.215&0.224&0.310&0.319&0.610&0.500&0.643 \\\hline
			$B (GeV^{-2})$	&16.5&18.3&18.5&19.1&21.3&21.6&33.9&35.2&36.5 \\\hline
		\end{tabular}%
	}
	\caption{Predictions using the unitarization procedure with $g_2(s)$.}
	\label{Table.4}
\end{table}

\section{Conclusions}
The observed non-monotonic rise of the slope $B(s,t)$ at the LHC is incompatible with a single pomeron pole. The odderon may play and important role in the behavior of $B(s)$ at high energies, as shown in Fig. \ref{Fig:Bo}. Even the combination of a simple and double pole (DP) cannot accommodate for the accelerated rise of $B(s)$ at the LHC, requiring unitarity corrections. At the same time, these data indicate the importance of unitarity corrections (see also \cite{Sch}) and point to the asymptotic regime, around 100 TeV.

To summarize:

a) The cross sections rise indefinitely, accelerating from $\ln s$ to $\ln^2 s$. The $\ln^2 s$ component is negligible in the energy regions of the ISR, making about $10\%$ of the cross sections, and starts to dominate in LHC energy region.

b) Upon unitarization, the onset of the asymptotics is determined by $g$ in
Eq.~(\ref{Eq:ge}). 

c) Our value of $\delta_P=0.0458$ in the case of the DP is about half of that in the case of a simple pole.

d) Diffractive scattering contains two components. One is universal, dominating at super-high energies, the second is reaction-dependent and can be seen slightly above the ISR energy region.     

e) It should be remembered that the analytic calculations in Secs.~\ref{Sec:Unitarity} and \ref{Sec:Asympt} were possible only approximately, up to $O(1/L)$. Precise results, including those in the non-foward direction can be performed numerically.

The unorthodox behaviour $\frac{\sigma_{el}}{\sigma_{tot}}\rightarrow 1$ as $s\rightarrow \infty$ is typical of the $u$-matrix approach \cite{TT}. It implies that the scattering hadrons, after crossing the black disc limit $\frac{\sigma_{el}}{\sigma_{tot}}=1/2$ (well beyond the LHC energy region!) will become more transparent. Accorting to Troshin and Tyurin \cite{TT}, it corresponds to the transition from shadow to antishadow scattering, called "reflective scattering".

\subsection*{Acknowledgements}

L. J. was supported by the Ukrainian Academy of Sciences' program "Nuclear matter under extreme conditions".

N. B. and I. Sz. thank to the organizers of the WE-Heraeus Physics School on QCD in Bad Honnef, where part of this work was done, for their hospitality and support. I. Sz. acknowledges also the hospitality and support at the BGL-10 conference in Gy\"ongy\"os.

\section*{Appendix}

The slope $B(s)$, calculated from Eq.~(\ref{Eq:slope}) with the norm Eq.~(\ref{norm}) and the amplitude Eq.~(\ref{Eq:Amplitude}) takes the form:
\begin{equation}
B(s)=\frac{a(s)+b(s)L+c(s)L^2+d(s)L^3}{e(s)+f(s)L+g(s)L^2},
\end{equation}
where
\begin{align} \label{Eq:abcdi}
a(s)&=\sum_{i=1}^{10}a_is^{k_i}, \ \ \ b(s)=\sum_{i=1}^{10}b_is^{k_i}, \ \ \ c(s)=\sum_{i=4}^{10}c_is^{k_i}, \ \ \ d(s)=\sum_{i=8}^{10}d_is^{k_i}, 
\\ \nonumber
& e(s)=\sum_{i=1}^{10}e_is^{k_i}, \ \ \ f(s)=\sum_{i=4}^{10}f_is^{k_i}, \ \ \ g(s)=\sum_{i=8}^{10}g_is^{k_i}.
\\ \nonumber
\end{align}
The parameters $a_i$, $b_i$, $c_i$, $d_i$, $e_i$, $f_i$, $g_i$, and $k_i$ are related to the paramters in Eqs.~(\ref{Reggeon1}-\ref{Eq:Otray}). By neglecting the oddereon, the terms where $i=5$, $7$, $9$ and $10$ will eliminate. Neglecting also the pomeron, terms with $i=4$, $6$ and $8$ disappear and we get
\begin{equation}
B(s)=\frac{a(s)+b(s)L}{e(s)}
\end{equation}
with terms where $i=1$, $2$ and $3$.
The expression for slope Eq.~(\ref{Eq:slope}) with the single pomeron Eq.(\ref{GP}) with trajectory Eq.~(\ref{Ptray}) reduces to:
\begin{equation} \label{Eq:DPB}
B(s)=\frac{(4m_\pi\alpha_{1P}+\alpha_{2P})(a+bL+cL^2+dL^3)}{2m_\pi(e+fL+gL^2)},
\end{equation}
where the parameters $a$, $b$, $c$, $d$, $e$, $f$ are energy-independent. They may be easily expressed in terms of those Eq.(\ref{GP}):
\begin{align}\label{abcd}
a&=b_{P}e^{b_{P}\delta_{P}}[4b_{P}^{2}e^{b_{P}\delta_{P}}+\pi^{2}(e^{b_{P}\delta_{P}}-\epsilon_{P})], \ \ \ b=\pi^{2}(e^{b_{P}\delta_{P}}-\epsilon_{P})^{2}+4b_{P}^{2}e^{b_{P}\delta_{P}}(3e^{b_{P}\delta_{P}}-\epsilon_{P}),
\\ \nonumber
&
c=12b_{P}e^{b_{P}\delta_{P}}(e^{b_{P}\delta_{P}}-\epsilon_{P}), \ \ \ d=4(e^{b_{P}\delta_{P}}-\epsilon_{P})^{2}, \ \ \ e=4b_{P}^{2}e^{2b_{P}\delta_{P}}+\pi^{2}(e^{b_{P}\delta_{P}}-\epsilon_{P})^{2},
\\ \nonumber
&   f=8b_{P}e^{b_{P}\delta_{P}}(e^{b_{P}\delta_{P}}-\epsilon_{P}), \ \ \ g=d=4(e^{b_{P}\delta_{P}}-\epsilon_{P})^{2}.
\\ \nonumber
\end{align}

In a similar way, the parameters $a_i$, $b_i$, $c_i$, $d_i$, $e_i$, $f_i$, $g_i$, and $k_i$ in Eq.~(\ref{Eq:abcdi}) may be related to those of Eqs.~(\ref{Reggeon1}-\ref{Eq:Otray}) (more complicated than in Eq.~(\ref{abcd})).

Alternatively, the local slope $B(s,t)$ with unit pomeron intercept $\alpha(t=0)=1$ can be written as Ref. \cite{Kholod, JS}

\begin{equation} \label{eq:B}
B(s,t)=2\alpha'(t)[b+F(s,t)L],
\end{equation}
where 
\begin{equation}\label{eq:B1}
F(s,t)=1+\Phi'(\alpha)\Bigl[1+\frac{\pi^2\Phi(\alpha)}{4L}+\Phi(\alpha) L\Bigr]\Bigl[[1+\Phi(\alpha) L]^2+\frac{\pi^2}{4}\Phi^2(\alpha)\Bigr]^{-1}.
\end{equation}
Equivalently, using Eq.(\ref{GP}) with the trajectory Eq.~(\ref{Ptray}) we get for the slope  Eq.~(\ref{Eq:DPB}).

Note  that $F(s,t)\rightarrow 1$ as $s\rightarrow \infty.$ Surprisingly, the function $F(s)$ decreases rapidly at small values of $s$, thus affecting the slope at small energies. It is close to $1$ at high energies where the pomeron dominates.

\end{document}